\documentclass[%
 reprint,
 amsmath,amssymb,
 aps,
]{revtex4-1}

\usepackage{graphicx}
\usepackage{dcolumn}
\usepackage{bm}
\usepackage{latexsym,amsthm}
\usepackage{graphics,physics}
\usepackage{graphicx}
\usepackage{epsfig}
\usepackage{multirow}
\usepackage{amsmath, amssymb}
\usepackage{hhline}
\usepackage{hyperref}
\usepackage{comment}
\usepackage{enumitem}
\usepackage{amsfonts}
\usepackage{xcolor}
\usepackage{soul}
\usepackage[compat=1.1.0]{tikz-feynman}
\usepackage{subcaption}
\usepackage{hhline}
\usepackage{bm}
\usepackage{float}
\restylefloat{table}
\usepackage[margin=1in]{geometry}
\usepackage{setspace}
\begin{document}

\preprint{APS/123-QED}

\title{Supersymmetric $SU(5) \times U(1)_\chi$ and the Weak Gravity Conjecture}

\author{Abhishek Pal}
\author{Qaisar Shafi}%
\affiliation{%
 Bartol Research Institute, Department of Physics and Astronomy\\
 University of Delaware, Newark, DE 19716, USA
}%

\date{\today}

\begin{abstract}
The gauge symmetry $SU(5) \times U(1)_\chi$ is the unique maximal subgroup of SO(10) which retains manifest unification at $M_{GUT}$ of the Standard Model gauge couplings, especially if low scale supersymmetry is present. The spontaneous breaking of $U(1)_\chi$ at some intermediate scale leaves unbroken a $Z_2$ symmetry  which is precisely `matter' parity. This yields a stable supersymmetric dark matter particle as well as topologically stable  cosmic strings. Motivated by the weak gravity conjecture we impose unification of $SU(5)$ and $U(1)_\chi$ at an ultraviolet cutoff  $\Lambda \sim \alpha_\Lambda ^{1/2} M_{P} \approx 5 \times 10^{17}$ GeV, where $\alpha_\Lambda$ denotes the $SU(5)$ gauge coupling at $\Lambda$ and $M_P \approx 2.4 \times 10^{18}$ GeV is the reduced Planck Scale. The impact of dimension five operators suppressed by $\Lambda$ on gauge coupling unification, proton lifetime estimates and $b-\tau$ Yukawa unification  is discussed. In particular, the gauge boson mediated proton decay into $e^+\pi^0$ can lie within the $2-\sigma$ sensitivity of HyperKamiokande. We also discuss how the intermediate scale strings may survive inflation while the $SU(5)$ monopoles are inflated away. The unbroken $Z_2$ symmetry provides an intriguing link between dark matter, black holes carrying `quantum hair' and cosmic strings.



\end{abstract}

\maketitle


\section{\label{sec:intro}Introduction}

 Grand Unification based on symmetry groups such as $SU(4)_c \times SU(2)_L  \times SU(2)_R$ \cite{Pati:1974yy}, $SU(5)$ \cite{Georgi:1974sy}, $SO(10)$ \cite{Proceedings:1975hpa,Fritzsch:1974nn} and $E_6$ \cite{Gursey:1975ki, Achiman:1978vg,Shafi:1978gg} provide  compelling frameworks for new physics
beyond the Standard Model (SM). Unification of the SM gauge couplings is most straightforwardly realized within the $SU(5)$ gauge group with low scale supersymmetry \cite{Dimopoulos:1981yj}. However, a discrete $Z_2$ symmetry or `matter' parity is required in $SU(5)$ to eliminate rapid proton decay and obtain a plausible dark matter candidate in the form of a neutral lightest supersymmetric particle (LSP). Moreover, since neutrinos are massless in the $SU(5)$ framework the observed solar and atmospheric neutrino oscillations cannot be explained. Of course, one is free to include $SU(5)$ singlet right handed neutrinos to resolve this latter problem, but this may not be entirely satisfactory because no symmetry exists to prevent the right handed neutrinos from acquiring arbitrarily large masses.


 The embedding of $SU(5)$ in an $SO(10)$ (more precisely $\text{Spin} (10)$) framework nicely resolves the problem of neutrino masses.  The presence of $U(1)_\chi$  requires three right handed neutrinos, inherited from $SO(10)$, which help implement the seesaw mechanism and explain the solar and atmospheric neutrino data. Furthermore, the right handed neutrinos acquire masses only after spontaneous breaking of $U(1)_\chi$ at some appropriately high scale, where $U(1)_\chi$  coincides with $U(1)_{B-L}$ for SM singlet fields such as the right handed neutrino.

Another important aspect of the $SO(10)$ symmetry that is relevant here has to do with its center $Z_4$. It was shown in ref~\cite{Kibble:1982ae} that the spontaneous breaking of $SO(10)$ to $SU(3)_c \times U(1)_{em}$ using only single valued (tensor) representations leaves unbroken the $Z_2$ subgroup of its center $Z_4$. In a supersymmetric setting this $Z_2$ is precisely ‘matter’ parity mentioned earlier. Since $SO(10)$ is a rank five group, the question naturally arises: how does $SO(10)$ break to the SM ? In a non-supersymmetric setting one or more intermediate scales are frequently employed to implement, among other things, gauge coupling unification. In a supersymmetric setting that concerns us here, the $SO(10)$ symmetry can be broken in a single step to the Minimal Supersymmetric Standard Model (MSSM), keeping intact the $Z_2$ symmetry. However, the Higgs fields required to break $SO(10)$ to
MSSM$\times Z_2$ reside in such large representations that the model becomes non-perturbative above $M_{GUT} (\approx 10^{16} \text{ GeV})$, the unification scale of the MSSM gauge couplings.

 To overcome this problem, and also motivated by the weak gravity conjecture, we propose to work instead with the maximal subgroup $SU(5) \times U(1)_\chi$, or
$\chi SU(5)$ for short. This subgroup of $SO(10)$ manifestly preserves gauge coupling unification, and it’s $U(1)_\chi$ component carries in it the $Z_4$ center of $SO(10)$, such that $Z_2$ survives at the end. According to the weak gravity conjecture, there exists an ultraviolet cutoff scale $\Lambda$ which, in the case of GUTs, is around $5 \times 10^{17}$ GeV. In the scenario we are proposing, the merger of $SU(5)$ and $U(1)_{\chi}$ gauge couplings occurs at a scale of order $\Lambda$. Between $M_{GUT}$ and $\Lambda$, the $\chi SU(5)$ theory remains fully perturbative. Furthermore, we can estimate how the presence of $\Lambda$ can impact some of the $SU(5)$ predictions including proton decay and $b-\tau$ Yukawa Unification. [Note that $t-b-\tau$ Yukawa Unification in $SO(10)$ \cite{Ananthanarayan:1991xp} may be realized at scale $\Lambda$.]

 The scale of $U(1)_\chi$ breaking can be estimated from a variety of considerations such as neutrino oscillations, inflation, leptogenesis and cosmic strings. 
We briefly discuss how these cosmic strings may survive inflation while the $SU(5)$ monopoles are inflated away.
 
\section{Gauge Coupling Unification and Weak Gravity Conjecture\label{sec:GCU}}

The field content of $\chi $SU(5) is displayed in Table ~\ref{tab:gauge_charges}. The matter multiplets come from three 16-plets of $SO(10)$ which contain the right handed neutrinos. In the Higgs sector we have the usual $24$-plet and also $5$, $\bar{5}$ fields which contain the two MSSM Higgs doublets. The $SU(5)$ singlet pair $\chi$, $\bar{\chi}$ acquire intermediate scale VEVs such that $U(1)_\chi$ is broken to $Z_2$, which is ‘matter’ parity. Note that the charge assignments listed in Table \ref{tab:gauge_charges} may suggest that the $\chi$, $\bar{\chi}$ VEVs break $U(1)_\chi$ to $Z_{10}$. However, since the $Z_5$ subgroup of $Z_{10}$ also resides in $SU(5)$, the effective unbroken discrete symmetry is $Z_2$.

\begin{table}[]
\begin{tabular}{|c|c|c|c|lll}
\cline{1-4}
\multicolumn{1}{|l|}{Group} & \multicolumn{3}{l|}{Representations}                                                                     &                                &                                      &                             \\ \cline{1-4}
\multicolumn{1}{|l|}{}      & \multicolumn{3}{c|}{Matter}                                                                              &                                &                                      &                             \\ \cline{1-4}
$SU(5)$                     & \multicolumn{1}{l|}{$F_i(\bar{5})$} & \multicolumn{1}{l|}{$T_i(10)$} & \multicolumn{1}{l|}{$\nu_i^c(1)$} &                                &                                      &                             \\ \cline{1-4}
$2\sqrt{10}U(1)_\chi$       & 3                                   & -1                             & -5                                &                                &                                      &                             \\ \hline
& \multicolumn{6}{|l|}{\hspace{0.85cm} Scalars}                                                                                                                                                                                                                \\ \hline
$SU(5)$                     & $\Phi(24)$                          & $H(5)$                         & $\bar{H}(\bar{5})$                & \multicolumn{1}{l|}{$\chi(1)$} & \multicolumn{1}{c|}{$\bar{\chi}(1)$} & \multicolumn{1}{l|}{$S(1)$} \\ \hline
$2\sqrt{10}U(1)_\chi$       & 0                                   & 2                              & -2                                & \multicolumn{1}{c|}{10}        & \multicolumn{1}{c|}{-10}             & \multicolumn{1}{c|}{0}      \\ \hline
\end{tabular}
\caption{Matter and Higgs content in minimal $SU(5) \times U(1)_\chi$. $\chi, \bar{\chi}$ fields implement $U(1)_\chi$  breaking and $\bar{\chi}$ provides masses to the right handed neutrinos, $\nu_i^c$. The singlet $S$ plays an important role during inflation.}
\label{tab:gauge_charges}
\end{table}

\begin{figure}[ht]
    \begin{subfigure}{0.5\textwidth}
    \includegraphics[scale=0.4]{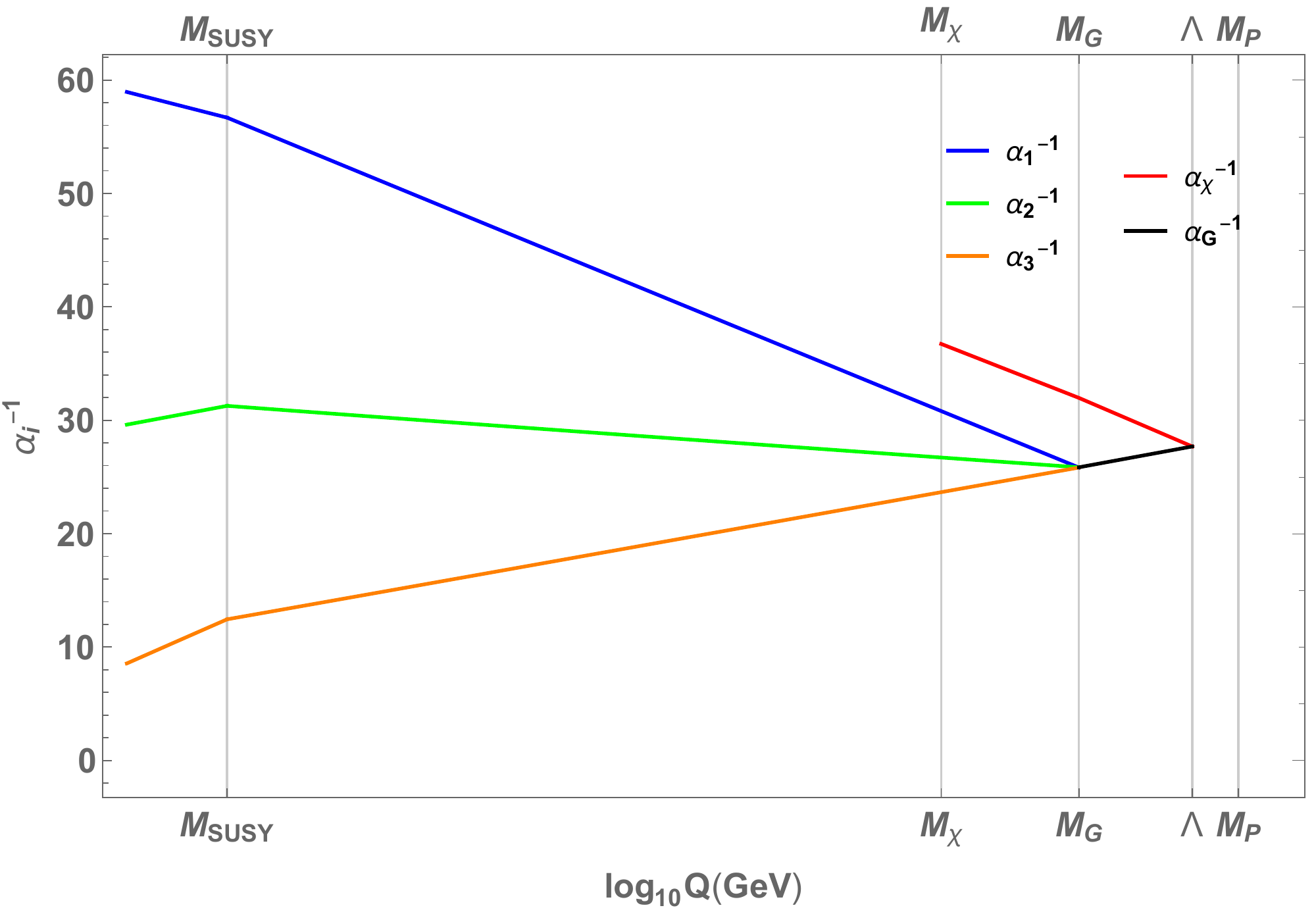}
    \end{subfigure}
    
    \caption{Running of gauge couplings in MSSM and $\chi SU(5)$. Unification of the $\chi SU(5)$ gauge couplings occurs at $\Lambda \approx 5 \times 10^{17}$GeV. $\mu_\chi = 10^{14}$GeV denotes the $U(1)_\chi$ symmetry breaking scale and $M_P = 2.4 \times 10^{18}$GeV is the reduced Planck scale.}
    \label{fig:Gauge running 1e14}
\end{figure}

\begin{figure}[ht]
    \begin{subfigure}{0.5\textwidth}
    \includegraphics[scale=0.4]{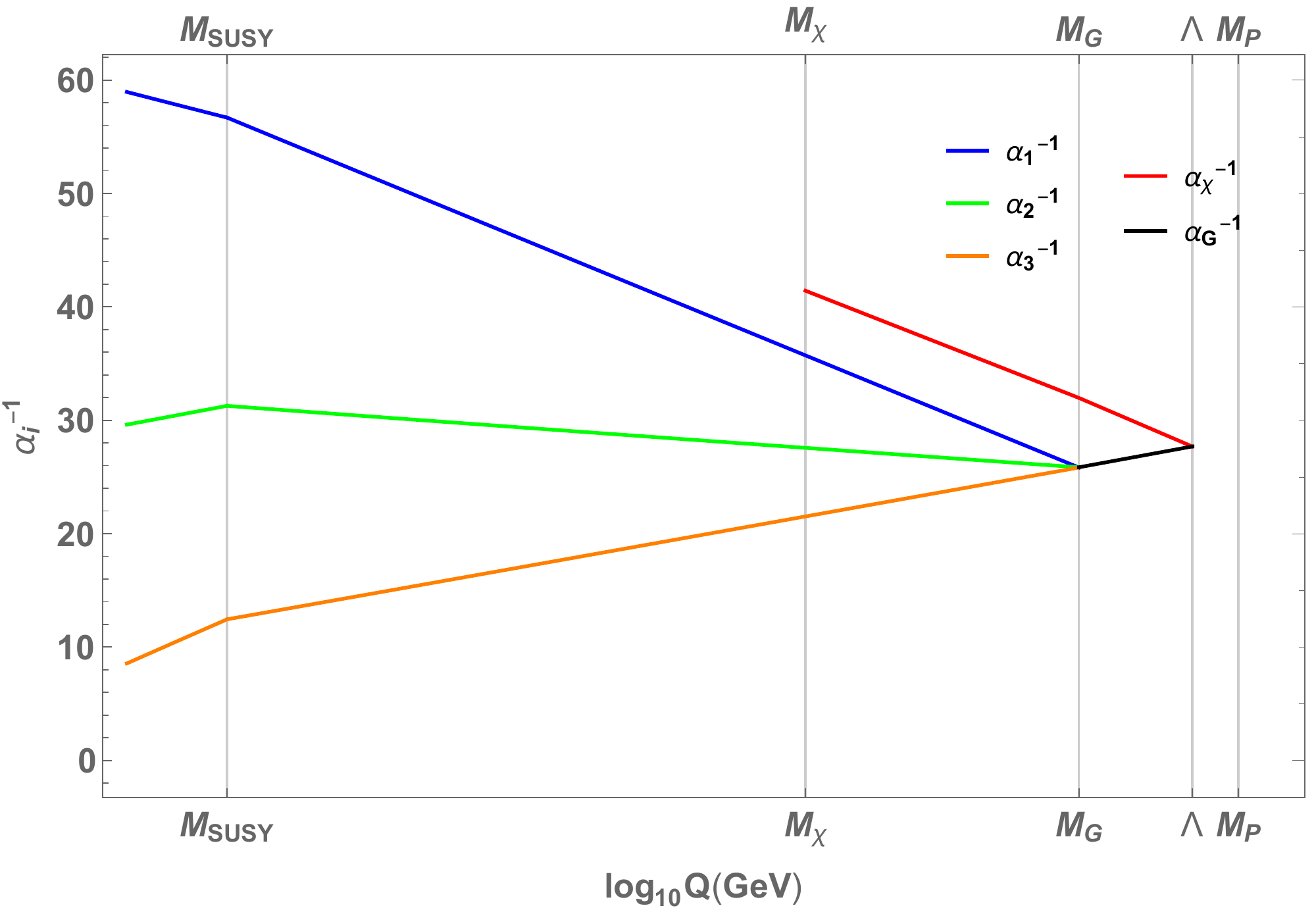}
    \end{subfigure}
    
    \caption{Same as FIG. 1 with  $\mu_\chi = 10^{12}$ GeV.}
    \label{fig:Gauge running 1e12}
\end{figure}

In Figures~[\ref{fig:Gauge running 1e14},\ref{fig:Gauge running 1e12}] we display unification of the MSSM gauge couplings at two loops using the software code SARAH \cite{Staub:2008uz}. The SUSY scale is taken to be $M_{SUSY} = \sqrt{m_{\Tilde{t}_1}m_{\Tilde{t}_2}} \approx 3$ TeV, where $m_{\Tilde{t}_1}$  and  $m_{\Tilde{t}_2}$ denote the stop masses. As expected, unification of the MSSM gauge couplings occurs at $M_{GUT} \approx 1.07 \times 10^{16}$ GeV. 

The figures also display unification of $\alpha_G$ and $\alpha_\chi$ which we assume occurs at the ultraviolet cutoff scale $\Lambda \approx \sqrt{\alpha_G} M_P$. 
The existence of $\Lambda$ lying between $M_{GUT}$ and $M_P$ is predicted by the weak gravity conjecture 
which is based on a variety of considerations including black holes and the non-existence of global symmetries in string theory~\cite{ArkaniHamed:2006dz}. In the context of  $\chi SU(5)$, this conjecture predicts $\Lambda \sim M_P \times \alpha_\Lambda^{1/2}$, where $\alpha_\Lambda$ denotes the unified coupling at scale $\Lambda$. In our case, $\Lambda$ turns out to be around $5 \times 10^{17}$ GeV if we identify $\alpha_\Lambda$ with  the unified coupling $\alpha_{G} \approx 1/25$. A comparable estimate for $\Lambda$ arises by noting that the effective field theory based on  $\chi SU(5)$ is asymptotically free and  viable  at energies $E > M_{GUT}$, as long as $\alpha _G$ stays larger than the dimensionless parameter $E^2 / M_P^2$ of gravity.

Note that an effective UV cutoff based on black hole physics and comparable to $\Lambda$, is given by $\Lambda_G \approx \frac{M_P}{\sqrt{N}}$, where $N$ denotes the number of particle species at scale $\Lambda_G$ \cite{Dvali:2007hz}.

In Figures \ref{fig:Gauge running 1e14} and \ref{fig:Gauge running 1e12}, the unified gauge coupling $\alpha_G$ is asymptotically free between $M_{GUT}$ and $\Lambda$. This wont be the case for $SO(10)$ running between $M_{GUT}$ and $\Lambda$.

 The appearance of the scale $\Lambda$ fairly close to $M_{GUT}$ suggests a possibly more significant role for higher dimensional operators in GUT related physics. In particular, dimension five operators [for earlier work see ref. \cite{Hill:1983xh, Shafi:1983gz,Calmet:2008df}] suppressed by a single power of $\Lambda$ could alter the predictions for $M_{GUT}$ which, in turn, would modify the standard proton lifetime predictions. In \cite{Shafi:1983gz} the scale $\Lambda$ was identified with the compactification scale of an underlying higher dimensional theory.
 
\begin{figure}
   \includegraphics[scale=0.38]{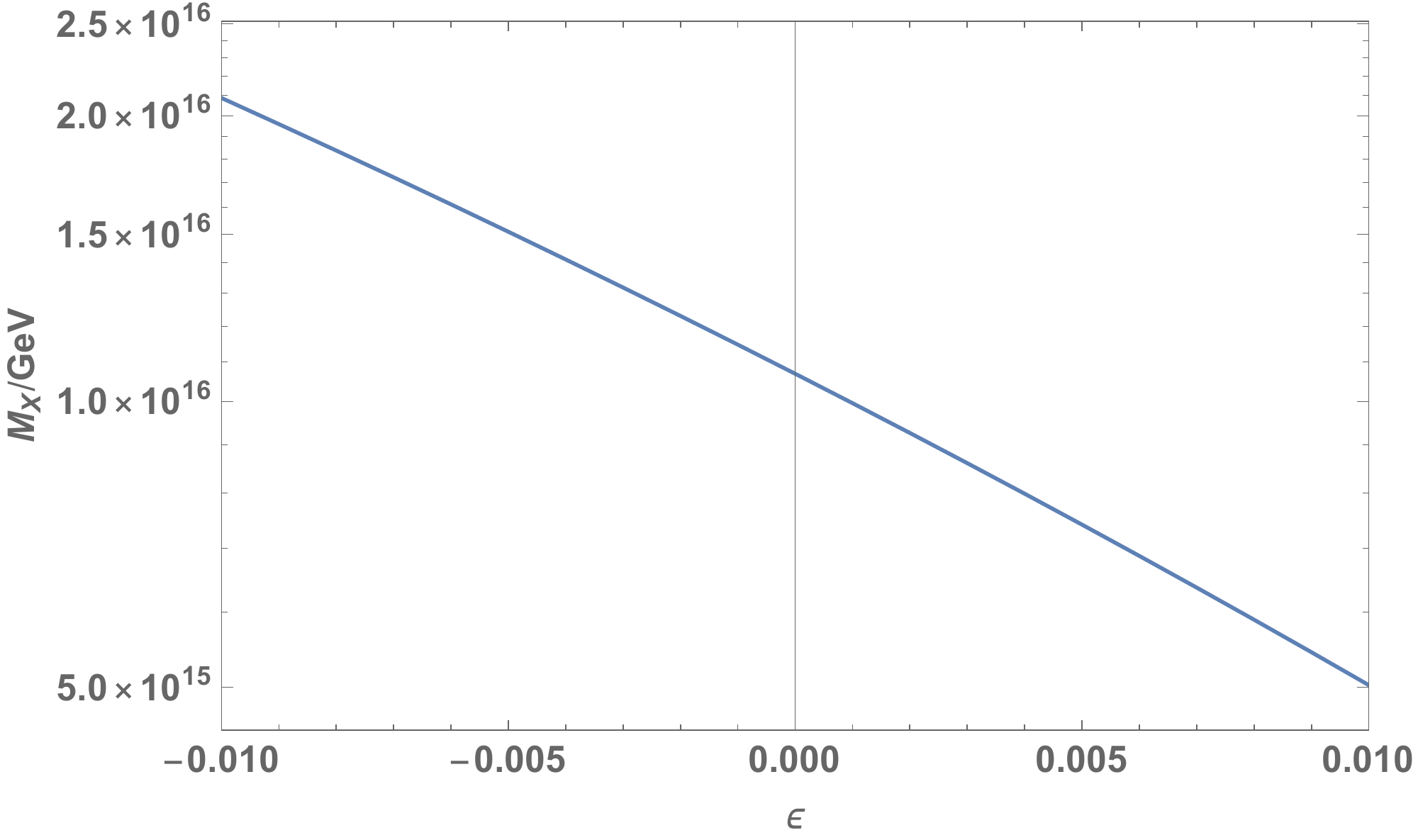}
    \caption{$M_\text{X}$(defined in eq.(\ref{eq:gauge boundary condition})) vs. $\epsilon$}
    \label{fig:Mx}
\end{figure}

\begin{figure}
\includegraphics[scale=0.38]{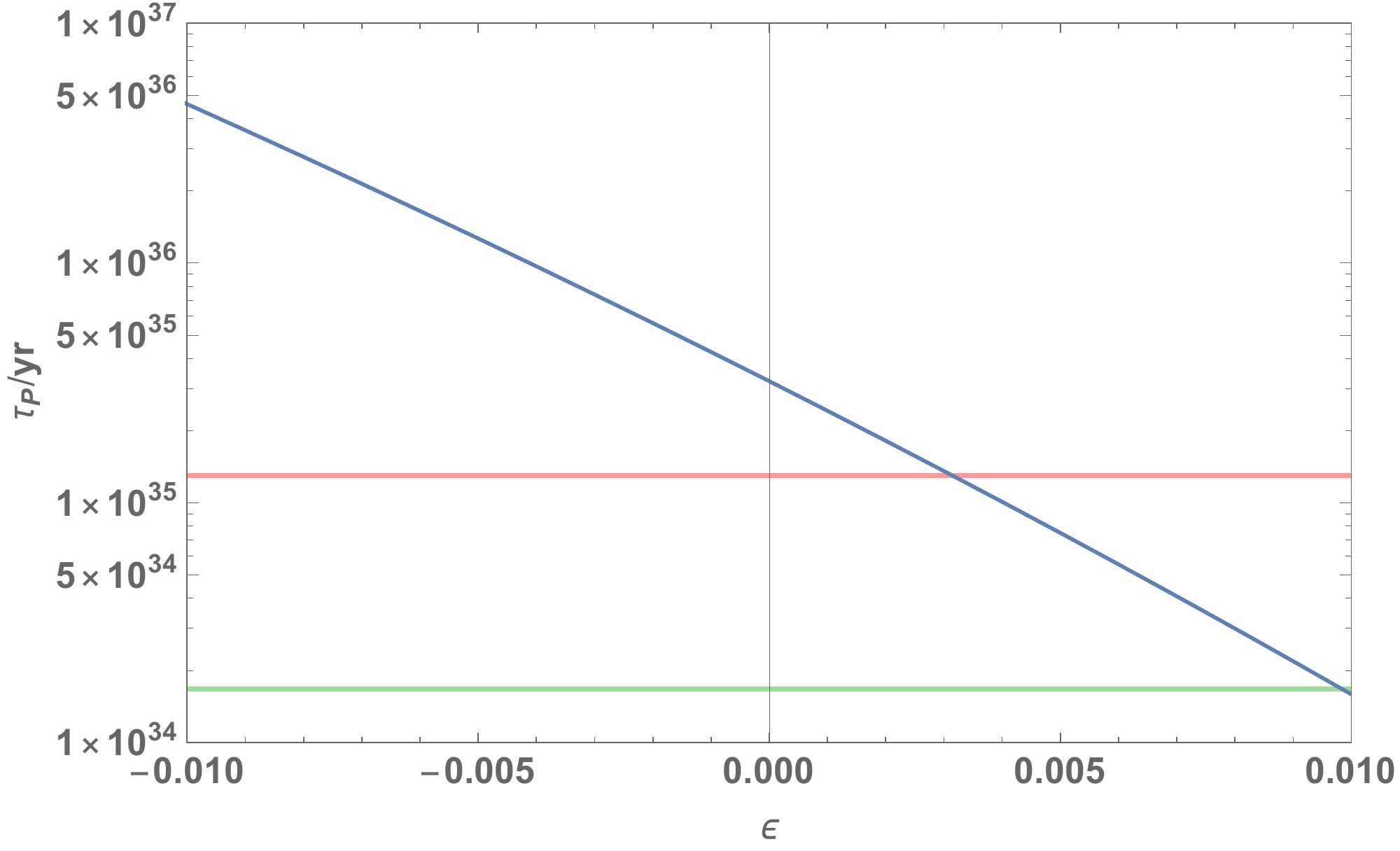}
    \caption{Proton Lifetime vs. $\epsilon$(blue line). The green line denotes the $2 \sigma$ experimental bound on proton lifetime set by Super-K\cite{Miura:2016krn}, and the red line is the expected $2\sigma$ sensitivity at Hyper-K \cite{Abe:2011ts}.}
    \label{fig:proton lifetime}
\end{figure}

Consider the dimension five term $\frac{\eta}{\Lambda}\text{ Tr}(F \cdot F \Phi$), where $\eta$ is a dimensionless constant, suppressed by the cutoff scale $\Lambda$. As shown in refs. \cite{Hill:1983xh, Shafi:1983gz} the unification conditions on the gauge couplings are modified as follows,
\begin{equation}
\begin{aligned}
(1+\epsilon)^{1/2}g_1(M_X) = (1+6 \epsilon)^{1/2}g_2(M_X)\\ = (1-4\epsilon)^{1/2}g_3(M_X).
\end{aligned}
\label{eq:gauge boundary condition}
\end{equation}
Here the dimensionless parameter $\epsilon \equiv \frac{\eta v}{\sqrt{15} \Lambda}$, where $v$ is the VEV of the $SU(5)$ adjoint Higgs multiplet, and $M_X$ plays the role of $M_{GUT}$ and coincides with it for $\epsilon  = 0$. In FIG. \ref{fig:Mx} we show a plot of $M_X$ versus $\epsilon$, and FIG.~\ref{fig:proton lifetime} shows the corresponding proton lifetime for $p \to e^+\pi^0$, which has been calculated using eq.(1.2) from \cite{Babu:2013jba},
\begin{equation}
\begin{aligned}
   \Gamma^{-1}(p \to e^+\pi^0) =  (1.6 \times 10^{35} \text{ yr}) \times \left( \frac{\alpha_H}{0.012 \text{ GeV}^3}\right)^{-2}\\
   \hspace{-2cm}\left( \frac{\alpha_G}{1/25}\right)^{-2}\left( \frac{A_R}{2.5}\right)^{-2}\left( \frac{M_X}{10^{16} \text{ GeV}}\right)^4.
   \end{aligned}
\end{equation}
Here $\alpha_H \simeq 0.01$ is the nuclear matrix element relevant for proton decay, and $A_R \simeq 2.5$ is the renormalization factor of the $d = 6$ proton decay operator. 

A suitably small positive value of $\epsilon$ shifts $M_X$ to lower values such that the proton lifetime lies within the range accessible by the Hyper-Kamiokande experiment. 

Regarding proton decay via Higgsino mediated dimension five operators, we assume that the SUSY scalars participating in this process are sufficiently heavy $( \gtrsim 20 \text{ TeV})$, such that the lifetime predictions do not contradict the current experimental bounds.

\section{$b-\tau$ Yukawa Unification\label{sec:yukawa}}
Many realistic $SU(5)$ models predict $b-\tau$ Yukawa Unification (YU) \cite{Ellis:1979fg} which would also hold for the $\chi$SU(5) model. Referring to Table \ref{tab:gauge_charges}, consider the following dimension-five terms that generate masses in SU(5) for down quarks and charged leptons \cite{Panagiotakopoulos:1984wf,Wiesenfeldt:2005zx, Calmet:2011ic}
\begin{equation}
    \begin{aligned}
    \frac{\varepsilon_{\alpha \beta \mu \nu \delta}}{\Lambda}\left( f_{ij} F_i^{\alpha \beta} T_j^{\mu \nu}\Phi^\delta_\rho\bar{H}^\rho + f'_{ij} F_i^{\alpha \beta} T_j^{\mu \rho}\bar{H}^\nu\Phi^\delta_\rho\right) + h.c.,
    \end{aligned}
    \label{eq:yukawa dimension 5}
\end{equation}
where $f_{ij}, f'_{ij}$ are dimensionless constants and the Greek letters denote the SU(5) indices. Ignoring the first two families, the usual $b-\tau$ Yukawa Unification condition at $M_{GUT}$ is modified to \cite{Wiesenfeldt:2005zx} 
\begin{equation}
    y_b-y_\tau \approx 5 f'_{33} \frac{M_{GUT}}{\Lambda}.
    \label{eq: yukawa MGUT}
\end{equation}
With $M_{GUT} \approx 2 \times 10^{16}$ GeV, $\Lambda \approx 5 \times 10^{17}$  GeV and $f'_{33}$ of order unity, this can modify exact $b-\tau$ YU by up to $10\%$ or so. This is in addition to any modifications arising from mixing between the families.

\begin{figure}[ht]
    \centering
    \includegraphics[scale=0.4]{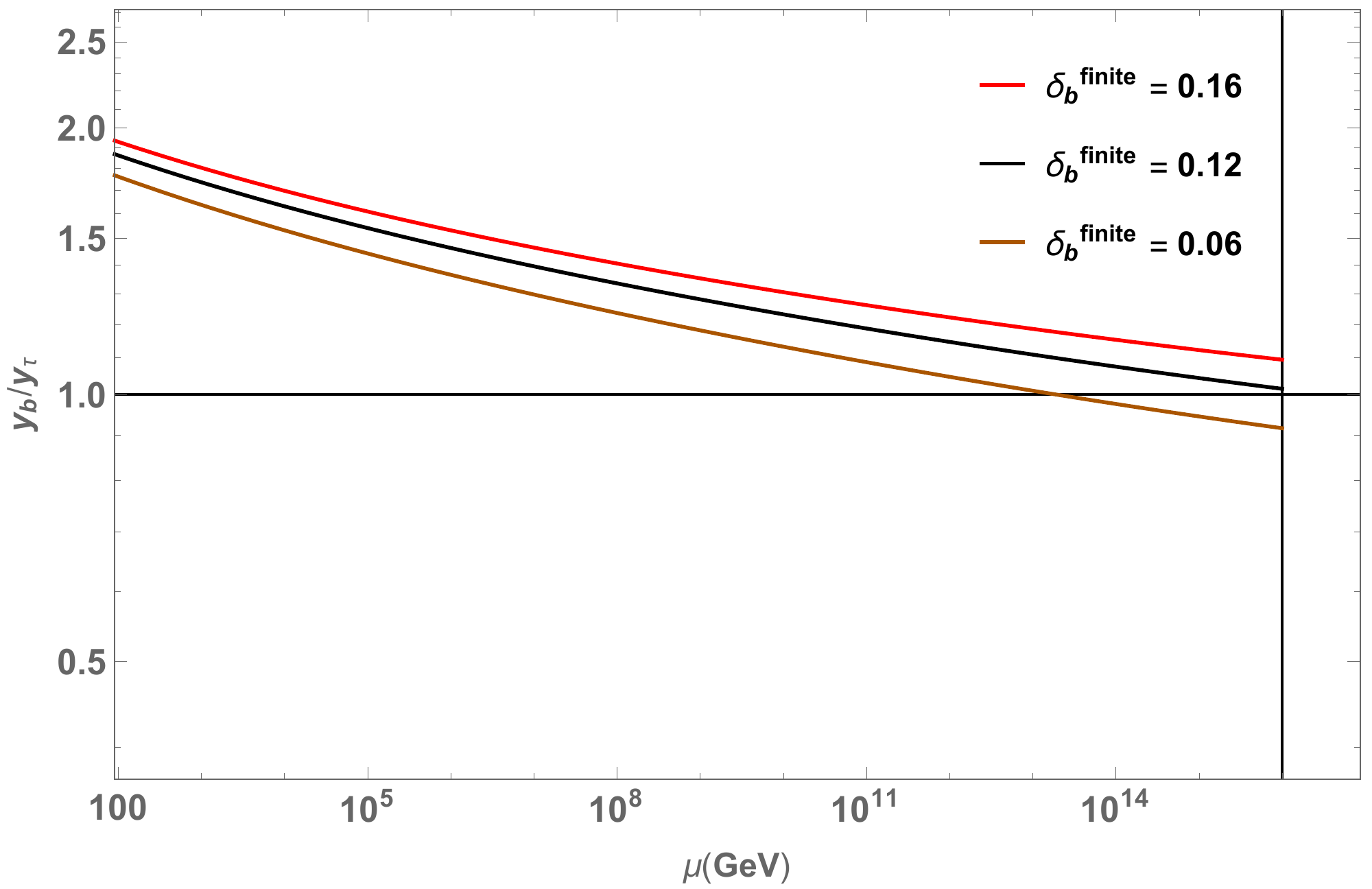}
    \caption{$y_b/y_\tau$ versus $\mu $, the energy scale, for tan$\beta = 50$. $y_b-y_\tau$ at $M_{GUT}$ are $0.06$ (top), $0$ (middle) and $-0.04$ (bottom). $\delta_b^{finite}$ denote the size of the finite one loop corrections to $y_b$. }
    \label{fig:yukawa tan beta=50}
\end{figure}

\begin{figure}[ht]
    \centering
    \includegraphics[scale=0.4]{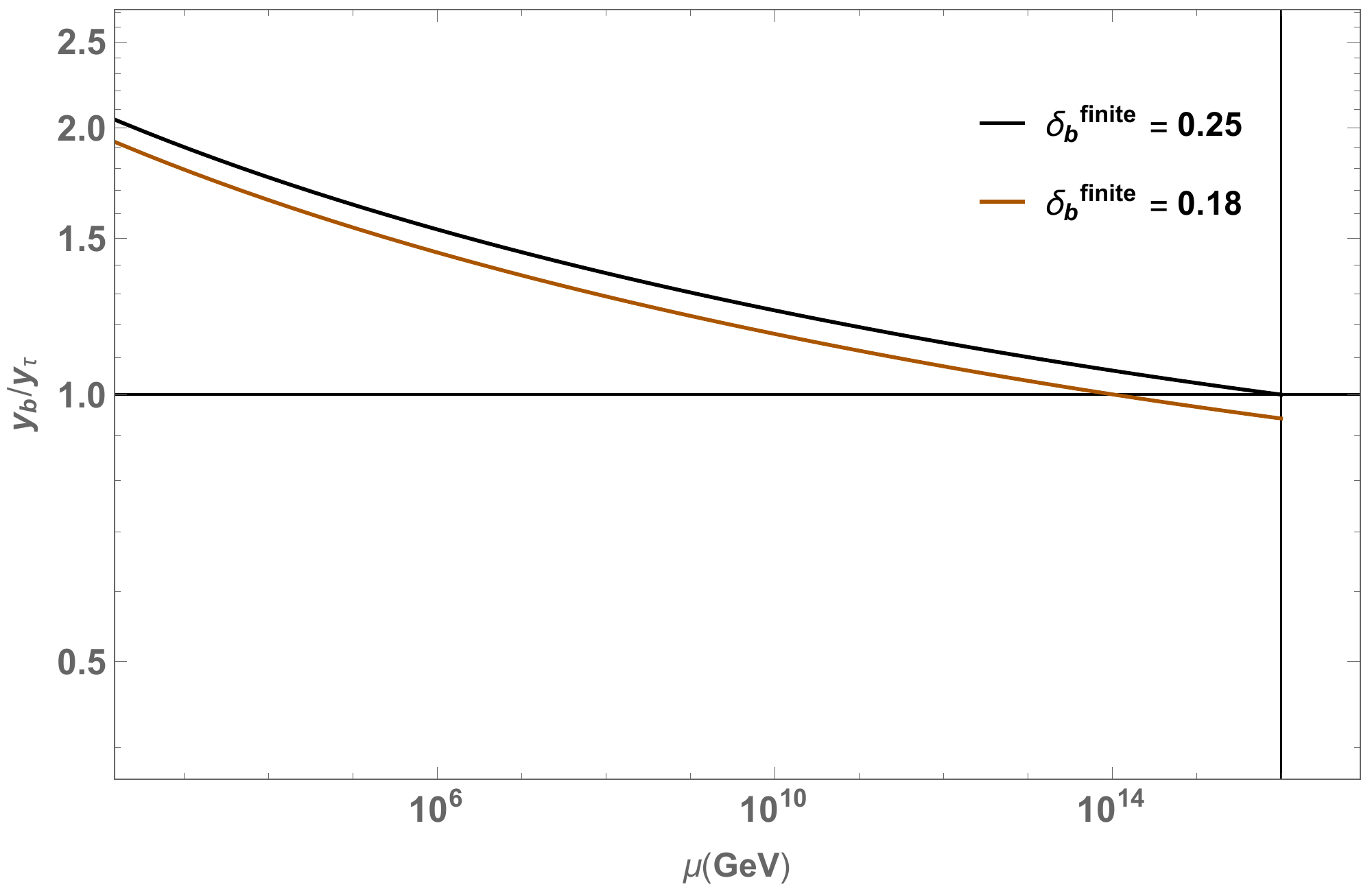}
    \caption{$y_b/y_\tau$ versus $\mu$, the energy scale, for tan$ \beta = 20$. $y_b-y_\tau$ at $M_{GUT}$ are $0$ (top curve) and $-0.01$ (bottom curve). $\delta_b^{finite}$ denote the size of the finite one loop corrections to $y_b$.}  
    \label{fig:yukawa tan beta=20}
\end{figure}

Finite one loop SUSY threshold corrections \cite{Hall:1993gn} are known to play an essential role in realizing $b-\tau$ YU in SUSY GUTs. In Figure ~\ref{fig:yukawa tan beta=50} we show the two loop running of $y_b$ and $y_\tau$  with tan $\beta = 50$, the SUSY scale is 3 TeV, and the leading radiative corrections to $y_b$, denoted by $\delta_b^{finite}$, \cite{Hall:1993gn, Elor:2012ig} vary from 6-16 \%. Radiative corrections to $y_t$ and $y_\tau$ will be ignored. 
For the corrections to $y_b$ set equal to 12 \% the YU condition is well satisfied, in agreement with the results in ref. \cite{Elor:2012ig}. However, deviations from this value yield approximate YU, which can be attributed to the presence of SU(5) breaking terms arising from the dimension five couplings in Eq.~(\ref{eq:yukawa dimension 5}). In Figure \ref{fig:yukawa tan beta=20} we display this phenomenon with tan $\beta$ set equal to 20.

A recent paper on $b-\tau$ YU in $SU(5)$ presented results based on a SUSY breaking scenario that yields non-universal gaugino masses at $M_{GUT}$ \cite{Raza:2018jnh}. A number of solutions compatible with the current experimental constraints from LHC, Planck and direct dark matter detection were found. These include gluino co-annihilation in which the gluino is nearly degenerate in mass ($\sim 1-2$ TeV) with the LSP neutralino, as well as several other cases in which the gluino can be considerably heavier,  of order 4 TeV or so. The benchmark points shown, which take into account the finite one loop corrections, exhibit $b-\tau$ Yukawa Unification at the level of $8-10 \%$. This can now be understood in light of the modified Yukawa condition in Eq.(\ref{eq: yukawa MGUT}).

\section{Inflation, Monopoles and Cosmic Strings in $SU(5) \times U(1)_\chi$\label{sec:inflation}}

To see how the cosmic strings may survive inflation while the monopoles are inflated away, consider the well-known superpotential $W$ and K\"ahler potential $K$ for implementing hybrid inflation associated with a symmetry breaking $G$ to $H$ \cite{Dvali:1994ms,Rehman:2009nq,Pallis:2013dxa,Buchmuller:2014epa},
\begin{equation}
\begin{aligned}
    W = {} & \kappa S (\Phi \bar{\Phi} - M^2),\\
    K = {} & |S|^2 + |\Phi|^2+|\bar{\Phi}|^2,
\end{aligned}
\label{eq: U1 W and K}    
\end{equation}
where $\Phi, \bar{\Phi}$ denote the conjugate pair of Higgs superfields responsible for breaking $G \to H$, $S$ is a gauge singlet field and $M$ denotes the scale at which $G$ is broken. A $U(1)$ R-symmetry restricts the renormalizable terms allowed in $W$. With minimal $W$ and $K$ this yields successful hybrid inflation in agreement with the Planck observations \cite{Aghanim:2018eyx}.

Inflation is driven by a scalar component of $S$, and $\Phi$, $\bar{\Phi}$, referred to as `waterfall' fields, acquire their VEVs at the end of inflation. If $G$ is $U(1)_\chi$ then cosmic strings will appear at the end of inflation. Following \cite{Pallis:2013dxa} the $U(1)_\chi$ symmetry breaking scale $\mu_\chi$ in this case can be as low as $6 \times 10^{14}$ GeV or so, which yields $G \mu \sim 1.5 \times 10^{-8}$ for the dimensionless string tension, where $G$ denotes Newton's constant and $\mu \simeq 2\pi \mu_\chi^2$ \cite{Hindmarsh:2011qj}. This prediction of $G \mu$ is compatible with the Planck bound $G \mu < 3.7 \times 10^{-7}$  derived from constraints on the string contribution to the CMB power spectrum \cite{Ade:2013xla}.  

A modified version of this minimal scenario employs a non-minimal K\"ahler potential \cite{BasteroGil:2006cm} and the inflationary predictions are in agreement with the recent Planck results \cite{Aghanim:2018eyx}. If the symmetry breaking $G$ to $H$ produces monopoles, we can use a non-minimal W  and minimal  or non-minimal K. In this so-called shifted hybrid inflation \cite{Jeannerot:2000sv} both $S$ and the waterfall fields take part in inflation, such that the monopoles are inflated away. 

Shifted hybrid inflation was successfully implemented in $SU(5)$ in \cite{Khalil:2010cp}.  To include $U(1)_\chi$ in this scenario we can introduce in W the following terms
\begin{equation}
    W \supset \sigma_\chi S \chi \bar{\chi} + \lambda T (\chi \bar{\chi}- \mu_\chi^2)+ \zeta \bar{\chi}\nu_i^c \nu_j^c,
    \label{eq: chi inflation}
\end{equation}
where $\nu_i^c,\nu_j^c$ denote the right handed neutrino superfields, and the second term implements $U(1)_\chi$ breaking along the lines mentioned earlier. Also, in K we include the terms 
\begin{equation}
    K \supset \frac{\kappa_{ST}}{M_p^2} |S|^2|T|^2 + \frac{\kappa_{S\chi}}{M_p^2} |S|^2|\chi|^2+\frac{\kappa_{S\bar{\chi}}}{M_p^2} |S|^2|\bar{\chi}|^2,
    \label{eq: chi Kahler}
\end{equation}
such that $T$ and $\chi$, $\bar{\chi}$ fields stay at the origin during inflation with suitable choices for the dimensionless parameters in Eq.(\ref{eq: chi Kahler}), while $S$ and the $SU(5)$ adjoint field participate in shifted hybrid inflation. In this case the $SU(5)$ monopoles are inflated away and cosmic strings appear after inflation is over.  Note that the dimensionless string tension in this case can be significantly lower, depending on the $U(1)_\chi$ breakig scale. The inclusion of $U(1)_\chi$ has the added advantage that we can implement leptogenesis at the end of inflation which we will not discuss here. [ For a discussion on how cosmic strings can survive inflation in non-supersymmetric $SU(5) \times U(1)_X$, with $U(1)_X$ a global symmetry, see ref.~\cite{Shafi:1984tt}.]

\section{Summary}
We have argued that $\chi SU(5)$, based on the gauge symmetry $SU(5) \times U(1)_\chi$, is a compelling
extension of the SM and MSSM, which presumably merges into $SO(10)$ and quantum gravity at a cutoff scale
$\Lambda \sim 5 \times 10^{17}$ GeV arising from the weak gravity conjecture. The $U(1)_\chi$ symmetry prevents rapid proton decay and its unbroken $Z_2$ subgroup ensures stability of the neutralino LSP, a viable dark matter candidate. With $M_X$ relatively close to $\Lambda$, we have explored its impact on gauge coupling unification, $b-\tau$ Yukawa unification and proton decay that arise from considerations of dimension five operators suppressed by $\Lambda$. We briefly discussed how a successful inflationary scenario can be realized in this framework such that the superheavy $SU(5)$ monopoles are inflated away but topologically stable cosmic strings from the intermediate scale breaking of $U(1)_\chi$ may be present in our galaxy. 

Finally, let us note that a black hole may carry a quantum number (`hair') \cite{ Coleman:1991sj} associated with the unbroken discrete $Z_2$ symmetry from $U(1)_\chi$, which suggests an intriguing relationship between black holes, dark matter and strings.

\section{Acknowledgment}
\normalfont Q.S thanks the DOE for partial support provided under grant number DE-SC 0013880.


\begin{thebibliography}{99}

\bibitem{Pati:1974yy} 
  J.~C.~Pati and A.~Salam,
  ``Lepton Number as the Fourth Color,''
  Phys.\ Rev.\ D {\bf 10}, 275 (1974)
  Erratum: [Phys.\ Rev.\ D {\bf 11}, 703 (1975)].
  
  \bibitem{Georgi:1974sy} 
  H.~Georgi and S.~L.~Glashow,
  ``Unity of All Elementary Particle Forces,''
  Phys.\ Rev.\ Lett.\  {\bf 32}, 438 (1974).

\bibitem{Proceedings:1975hpa} 
  H.~C.~C.~E.~W.~Carlson,
  ``PARTICLES AND FIELDS — 1974: Proceedings of the Williamsburg Meeting of APS/DPF,''
  AIP Conf.\ Proc.\  {\bf 23} (1975).
  
  \bibitem{Fritzsch:1974nn} 
  H.~Fritzsch and P.~Minkowski,
  ``Unified Interactions of Leptons and Hadrons,''
  Annals Phys.\  {\bf 93}, 193 (1975).
  
  
    \bibitem{Gursey:1975ki} 
  F.~Gursey, P.~Ramond and P.~Sikivie,
  ``A Universal Gauge Theory Model Based on E6,''
  Phys.\ Lett.\  {\bf 60B}, 177 (1976).
  
  \bibitem{Achiman:1978vg} 
  Y.~Achiman and B.~Stech,
  ``Quark Lepton Symmetry and Mass Scales in an E6 Unified Gauge Model,''
  Phys.\ Lett.\  {\bf 77B}, 389 (1978).
  
  \bibitem{Shafi:1978gg} 
  Q.~Shafi,
  ``E(6) as a Unifying Gauge Symmetry,''
  Phys.\ Lett.\  {\bf 79B}, 301 (1978).
  
  \bibitem{Dimopoulos:1981yj} 
  S.~Dimopoulos, S.~Raby and F.~Wilczek,
  Phys.\ Rev.\ D {\bf 24}, 1681 (1981).
  doi:10.1103/PhysRevD.24.1681
  
  
\bibitem{Kibble:1982ae} 
  T.~W.~B.~Kibble, G.~Lazarides and Q.~Shafi,
  ``Strings in SO(10),''
  Phys.\ Lett.\  {\bf 113B}, 237 (1982).

\bibitem{Ananthanarayan:1991xp} 
  B.~Ananthanarayan, G.~Lazarides and Q.~Shafi,
  Phys.\ Rev.\ D {\bf 44}, 1613 (1991).
  doi:10.1103/PhysRevD.44.1613
  
  
  \bibitem{Staub:2008uz} 
  F.~Staub,
  ``Sarah,''
  arXiv:0806.0538 [hep-ph].

\bibitem{ArkaniHamed:2006dz} 
  N.~Arkani-Hamed, L.~Motl, A.~Nicolis and C.~Vafa,
  ``The String landscape, black holes and gravity as the weakest force,''
  JHEP {\bf 0706}, 060 (2007)
  [hep-th/0601001].

  
    \bibitem{Dvali:2007hz} 
  G.~Dvali,
  ``Black Holes and Large N Species Solution to the Hierarchy Problem,''
  Fortsch.\ Phys.\  {\bf 58}, 528 (2010)
  [arXiv:0706.2050 [hep-th]];
  G.~Dvali and M.~Redi,
  ``Black Hole Bound on the Number of Species and Quantum Gravity at LHC,''
  Phys.\ Rev.\ D {\bf 77}, 045027 (2008)
  [arXiv:0710.4344 [hep-th]].
  
  \bibitem{Hill:1983xh} 
  C.~T.~Hill,
  ``Are There Significant Gravitational Corrections to the Unification Scale?,''
  Phys.\ Lett.\  {\bf 135B}, 47 (1984).
  
\bibitem{Shafi:1983gz} 
  Q.~Shafi and C.~Wetterich,
  ``Modification of {GUT} Predictions in the Presence of Spontaneous Compactification,''
  Phys.\ Rev.\ Lett.\  {\bf 52}, 875 (1984).
  
   \bibitem{Calmet:2008df} 
  X.~Calmet, S.~D.~H.~Hsu and D.~Reeb,
  ``Grand unification and enhanced quantum gravitational effects,''
  Phys.\ Rev.\ Lett.\  {\bf 101}, 171802 (2008)
  [arXiv:0805.0145 [hep-ph]].
  
  \bibitem{Babu:2013jba} 
  K.~S.~Babu {\it et al.},
  ``Working Group Report: Baryon Number Violation,''
  arXiv:1311.5285 [hep-ph].
  
  
    \bibitem{Miura:2016krn} 
  K.~Abe {\it et al.} [Super-Kamiokande Collaboration],
  ``Search for proton decay via $p \to e^+\pi^0$ and $p \to \mu^+\pi^0$ in 0.31  megaton·years exposure of the Super-Kamiokande water Cherenkov detector,''
  Phys.\ Rev.\ D {\bf 95}, no. 1, 012004 (2017)
  [arXiv:1610.03597 [hep-ex]].

  \bibitem{Abe:2011ts} 
  K.~Abe {\it et al.},
  ``Letter of Intent: The Hyper-Kamiokande Experiment --- Detector Design and Physics Potential ---,''
  arXiv:1109.3262 [hep-ex].
  
    

  
  
      \bibitem{Ellis:1979fg} 
  J.~R.~Ellis and M.~K.~Gaillard,
  ``Fermion Masses and Higgs Representations in SU(5),''
  Phys.\ Lett.\  {\bf 88B}, 315 (1979).
  
   \bibitem{Panagiotakopoulos:1984wf} 
  C.~Panagiotakopoulos and Q.~Shafi,
  ``Dimension Five Interactions, Fermion Masses and Higgs Mediated Proton Decay in SU(5),''
  Phys.\ Rev.\ Lett.\  {\bf 52}, 2336 (1984).
  
  \bibitem{Wiesenfeldt:2005zx} 
  S.~Wiesenfeldt,
  ``Operator analysis for proton decay in SUSY SO(10) GUT models,''
  Phys.\ Rev.\ D {\bf 71}, 075006 (2005)
  [hep-ph/0501223].

\bibitem{Calmet:2011ic} 
  X.~Calmet and T.~C.~Yang,
  ``Gravitational Corrections to Fermion Masses in Grand Unified Theories,''
  Phys.\ Rev.\ D {\bf 84}, 037701 (2011)
  [arXiv:1105.0424 [hep-ph]].
  
    \bibitem{Hall:1993gn} 
  L.~J.~Hall, R.~Rattazzi and U.~Sarid,
  ``The Top quark mass in supersymmetric SO(10) unification,''
  Phys.\ Rev.\ D {\bf 50}, 7048 (1994)
  [hep-ph/9306309].
  
  \bibitem{Elor:2012ig} 
  G.~Elor, L.~J.~Hall, D.~Pinner and J.~T.~Ruderman,
  ``Yukawa Unification and the Superpartner Mass Scale,''
  JHEP {\bf 1210}, 111 (2012)
  [arXiv:1206.5301 [hep-ph]].
  
   \bibitem{Raza:2018jnh} 
  S.~Raza, Q.~Shafi and C.~S.~Un,
  ``$b-\tau$ Yukawa Unification in SUSY SU(5) with Mirage Mediation: LHC and Dark Matter Implications,''
  arXiv:1812.10128 [hep-ph].
  
   \bibitem{Dvali:1994ms} 
  G.~R.~Dvali, Q.~Shafi and R.~K.~Schaefer,
  ``Large scale structure and supersymmetric inflation without fine tuning,''
  Phys.\ Rev.\ Lett.\  {\bf 73}, 1886 (1994)
  [hep-ph/9406319].
  
  \bibitem{Rehman:2009nq} 
  M.~U.~Rehman, Q.~Shafi and J.~R.~Wickman,
  ``Supersymmetric Hybrid Inflation Redux,''
  Phys.\ Lett.\ B {\bf 683}, 191 (2010)
  [arXiv:0908.3896 [hep-ph]].
  
    \bibitem{Pallis:2013dxa} 
  C.~Pallis and Q.~Shafi,
  ``Update on Minimal Supersymmetric Hybrid Inflation in Light of PLANCK,''
  Phys.\ Lett.\ B {\bf 725}, 327 (2013)
  [arXiv:1304.5202 [hep-ph]].



  
  
    \bibitem{Buchmuller:2014epa} 
  W.~Buchmüller, V.~Domcke, K.~Kamada and K.~Schmitz,
  ``Hybrid Inflation in the Complex Plane,''
  JCAP {\bf 1407}, 054 (2014)
  [arXiv:1404.1832 [hep-ph]].
  
    \bibitem{Aghanim:2018eyx} 
  N.~Aghanim {\it et al.} [Planck Collaboration],
  ``Planck 2018 results. VI. Cosmological parameters,''
  arXiv:1807.06209 [astro-ph.CO];
  
\bibitem{Hindmarsh:2011qj} 
  M.~Hindmarsh,
  ``Signals of Inflationary Models with Cosmic Strings,''
  Prog.\ Theor.\ Phys.\ Suppl.\  {\bf 190}, 197 (2011)
  doi:10.1143/PTPS.190.197
  [arXiv:1106.0391 [astro-ph.CO]].
  
  
  \bibitem{Ade:2013xla} 
  P.~A.~R.~Ade {\it et al.} [Planck Collaboration],
  ``Planck 2013 results. XXV. Searches for cosmic strings and other topological defects,''
  Astron.\ Astrophys.\  {\bf 571}, A25 (2014)
  doi:10.1051/0004-6361/201321621
  [arXiv:1303.5085 [astro-ph.CO]].
  
   \bibitem{BasteroGil:2006cm} 
  M.~Bastero-Gil, S.~F.~King and Q.~Shafi,
  ``Supersymmetric Hybrid Inflation with Non-Minimal Kahler potential,''
  Phys.\ Lett.\ B {\bf 651}, 345 (2007)
  [hep-ph/0604198].
 
 \bibitem{Jeannerot:2000sv} 
  R.~Jeannerot, S.~Khalil, G.~Lazarides and Q.~Shafi,
  ``Inflation and monopoles in supersymmetric SU(4)C x SU(2)(L) x SU(2)(R),''
  JHEP {\bf 0010}, 012 (2000)
  doi:10.1088/1126-6708/2000/10/012
  [hep-ph/0002151].



   \bibitem{Khalil:2010cp} 
  S.~Khalil, M.~U.~Rehman, Q.~Shafi and E.~A.~Zaakouk,
  ``Inflation in Supersymmetric SU(5),''
  Phys.\ Rev.\ D {\bf 83}, 063522 (2011)
  [arXiv:1010.3657 [hep-ph]].
  
  \bibitem{Shafi:1984tt} 
  Q.~Shafi and A.~Vilenkin,
  ``Spontaneously Broken Global Symmetries and Cosmology,''
  Phys.\ Rev.\ D {\bf 29}, 1870 (1984).


  \bibitem{Coleman:1991sj} 
  L.~M.~Krauss and F.~Wilczek,
  ``Discrete Gauge Symmetry in Continuum Theories,''
  Phys.\ Rev.\ Lett.\  {\bf 62}, 1221 (1989).
  
  S.~R.~Coleman, J.~Preskill and F.~Wilczek,
  ``Dynamical effect of quantum hair,''
  Mod.\ Phys.\ Lett.\ A {\bf 6}, 1631 (1991).
  
\end{thebibliography}
\end{document}